# Title

Improved cycling stability and lithium utilization in trilayer Al-LLZO revealed by Electrochemical cycling performance

# Authors

Naisargi Kanabar[1] ( nkanabar@albany.edu ) - Corresponding Author
Seiichiro Higashiya[2] ( shigashiya@albany.edu )
Haralabos Efstathiadis[2] ( hefstathiadis@albany.edu )

# Affiliations

[1]Department of Physics, University at Albany, Albany, NY 12222
[2]Department of Nanoscale Science and Engineering, University at Albany, Albany, NY 12222

# Abstract

Garnet-type $Li_{6.25}Al_{0.25}La_3Zr_2O_{12}$ (Al-LLZO) solid electrolytes are promising for all-solid-state batteries but are limited by interfacial resistance. In this work, dense and graded tri-layer Al-LLZO electrolytes were fabricated and tested in Li/Al-LLZO/NMC(111) full cells. After 25 cycles, the tri-layer cell delivered discharge capacity of ~55 mAh g$^{-1}$, nearly twice that of the dense Al-LLZO (~27 mAh g$^{-1}$). EIS showed lower initial interfacial resistance (~373 Ω) and improved stability. SEM confirmed a porous-dense-porous structure, while NRA revealed enhanced near surface lithium (~75%) compared to dense Al-LLZO (~48%). These results highlight the role of microstructural grading in improving lithium distribution and cell performance.

# Keywords

# 1. Introduction

Driven by the growing demand for high-energy-density electronics and electric vehicles, lithium metal, featuring an exceptional theoretical capacity (3860 mAh g$^{-1}$) and lowest reduction potential (−3.04 V vs. standard hydrogen electrode) has re-emerged as a promising anode for next-generation batteries [1,2]. However, the development of high-energy-density lithium-metal batteries is limited by lithium dendrite formation. A solid electrolyte can be a better option to mitigate this issue. Among all solid electrolytes, aluminum-doped lithium lanthanum zirconium oxide (Al-LLZO) garnet-type ceramics offer high room temperature (RT) ionic conductivity (~$10^{-4}$ S cm$^{-1}$), a wide electrochemical stability window (0-6 V vs Li/Li$^+$), and good chemical compatibility with Li metal, making them a promising choice for solid-state lithium-metal batteries [3,4]. Despite these advantages, the practical implementation of Al-LLZO based solid-state batteries remains challenging due to high interfacial resistance and poor physical contact between the rigid Al-LLZO electrolyte and the electrode materials. The use of a solid electrolyte next to the anode and a liquid electrolyte next to the cathode has been proposed by Goodenough as a strategy to enhance safety while maintaining high energy density [5].

To address these issues, two approaches can be considered. J. Wang et al. demonstrated that larger Li particles grown at elevated temperatures reduce the electrolyte/electrode interfacial area [6]. Therefore, applying heat treatment to the Li metal before cell assembly can improve its contact with the solid-state electrolyte. In addition, S.L. Beshahwured et al. showed that a tri-layer membrane can suppress Li dendrite growth in all-solid-state Li-metal batteries [7]. Designing Al-LLZO with a tri-layer structure can combine the mechanical robustness of dense layers with the enhanced ionic transport and flexibility of porous or modified intermediate layers, thereby reducing Li dendrite growth. Such multilayer architectures can also lower interfacial impedance, accommodate stress during cycling, and promote uniform lithium-ion transport.

In this work, we compare the electrochemical performance of cells employing dense Al-LLZO and tri-layer Al-LLZO electrolytes in Li/Al-LLZO/NMC(111) configurations. Electrochemical Impedance Spectroscopy (EIS), cycling performance, and voltage–capacity analysis were carried out to evaluate interfacial behavior and stability.

The results demonstrate that the tri-layer Al-LLZO exhibits significantly improved cycling stability, lower interfacial resistance, and higher capacity retention compared to the dense Al-LLZO, confirming the effectiveness of the tri-layer design in enhancing solid-state cell performance. Nuclear Reaction Analysis (NRA) further shows that the tri-layer Al-LLZO retains a higher lithium concentration after 25 cycles relative to dense Al-LLZO. This enhanced lithium retention is consistent with the improved electrochemical cycling stability, suggesting more stable lithium transport and reduced interfacial lithium loss in the tri-layer architecture. Overall, these results indicate that the tri-layer structure facilitates reversible lithium accommodation at the interface, thereby mitigating interfacial degradation during cycling.

# 2. Experimental

## 2.1. Tape casting of Al-LLZO

The synthesis of Al-LLZO pellets was adapted from the procedures reported by Yi et al., and Hitz et al. [8-9], involving a four-step process: slurry preparation, tape casting, tri-layer assembly, and sintering/calcination. Aluminum-doped LLZO, lithium carbonate, PVB, BBP, ethanol, and acetone were milled for 48 h with yttria-stabilized zirconia (YSZ) media to form a homogeneous slurry. For the porous layer, PMMA microspheres were incorporated as porogens and milled for an additional hour before tape casting onto a Mylar substrate and drying overnight. The dense and porous layers were subsequently laminated by hot pressing at 60 °C for 20 min to form a bilayer, followed by the addition of a second porous layer to construct the tri-layer host structure. The films were peeled from the substrate, cut into 22 mm discs, sintered at 1080 °C for 1 h under argon, and finally calcined at 840 °C in air to remove residual carbonaceous materials.

## 2.2. Cell assembly

Full NMC111/Li Li-ion cells were assembled using sintered Al-LLZO pellets. Before assembly, the Li chip (15.6 mm diameter x 0.45 mm thickness) was heated at 60 °C for 10 s. A liquid electrolyte (1 M $LiPF_6$, LP 572, Gotion, Fremont, CA) was added (100 μL) to the cathode side for wetting purposes. Cell assembly was carried out in an argon-filled glove box using a PAT-Cell design (EL-Cell GmbH, Hamburg, Germany) with insulation sleeve (PP), Li ring, FS/5P separator (ECC1-00-0210-V/X).

## 2.3. Full cell cycling test

Electrochemical cycling was performed using an Arbin BT2453 battery cycler. The cells were charged at a constant current of C/10 (0.52 mA) until reaching 4.1 V, followed by a constant-voltage step at 4.1 V until the current decreased to 0.13 mA. After charging, the cells were rested for 20 minutes. Discharge was carried out at a constant current of C/10 (−0.52 mA) down to 3.5 V, followed by another 20-minute rest period. A total of 25 charge–discharge cycles were completed for both dense and tri-layer Al-LLZO under these conditions. EIS measurements were conducted using a Gamry potentiostat over a frequency range from 0.1 Hz to 100 KHz with an AC perturbation amplitude of 10 mV.

## 2.4. Postmortem analysis of LLZO

The lithium content of tri-layer and dense Al-LLZO samples after 25 cycles was quantified using NRA based on the $^7$Li(p,α)$^4$He reaction [10]. This p-α NRA approach was selected due to its high sensitivity, minimal dependence on sample composition, and suitability for thick solid electrolyte samples, in contrast to proton-induced gamma-ray emission (PIGE), which is more strongly affected by matrix effects [11]. The samples were irradiated with 1.2 MeV protons generated by a Dynamitron accelerator (Ion Beam Lab, University at Albany), and the emitted α-particles were detected using a silicon (Si) detector at a scattering angle of 169.87°. To suppress the intense elastic backscattering of protons, a plastic absorber foil was placed in front of the Si detector. The total collected charge was 2 μC, the detector calibration was 4.11 keV per channel, and the experimental spectra were analyzed using SIMNRA software (version 7.03) to obtain quantitative lithium depth profiles.

The morphology and microstructure of the samples were examined using a Zeiss Leo 1550 scanning electron microscope (SEM).

# 3. Results and Discussion

## 3.1. Electrochemical cycling of the full cells

Figure 1 shows the cell voltage versus specific capacity profiles for Li/Al-LLZO/NMC(111) cells employing dense Al-LLZO (Fig. 1a) and tri-layer Al-LLZO garnets (Fig. 1b). Both cells were cycled 25 times at RT. After 25 cycles, the tri-layer Al-LLZO cell delivers a discharge capacity of 54.94 mAh g$^{-1}$, which is approximately twice that of the dense Al-LLZO cell (26.54 mAh g$^{-1}$). The superior electrochemical performance of the tri-layer Al-LLZO can be attributed to its improved interfacial contact and reduced interfacial resistance, which enhances lithium-ion transport during cycling. Additionally, the multilayer structure may help to accommodate mechanical stresses and mitigate interfacial degradation, leading to more stable cycling behavior compared to the dense Al-LLZO electrolyte.

## 3.2. EIS data

Figure 2a represents the tri-layer Al-LLZO structure, which exhibits improved coulombic efficiency (CE) relative to the dense Al-LLZO cell, indicating enhanced reversibility and interfacial stability during cycling.

Figure 2b illustrates the Nyquist plots of Li/Al-LLZO/NMC (111) cells using dense and tri-layer Al-LLZO separators, recorded before cycling and after 25 cycles. Before cycling, the dense Al-LLZO cell exhibited a high interfacial impedance of 992.2 Ω, which significantly decreased to 163.5 Ω after cycling, possibly due to improved interfacial contact during the initial cycles. In contrast, the tri-layer Al-LLZO cell showed a much lower initial impedance of 372.8 Ω, which slightly increased to 457.6 Ω after 25 cycles, indicating minor interfacial changes. Despite this small increase, the tri-layer Al-LLZO maintained lower overall resistance and more stable impedance behavior compared to the dense Al-LLZO, demonstrating its superior interfacial stability and improved ionic transport characteristics during repeated cycling.

## 3.3. SEM analysis

Figure 3a & b show top-view SEM images of dense and porous Al-LLZO electrolytes sintered at 1080 °C, respectively. The magnified image of dense Al-LLZO (Fig. 3c) reveals a highly compact microstructure with well-sintered grains and minimal porosity, indicating effective densification. In contrast, the magnified image of porous Al-LLZO (Fig. 3d) exhibits visible pores at the same magnification, confirming the presence of an interconnected porous structure.

Figure 3e & f present cross-sectional SEM images of the tri-layer Al-LLZO membrane before and after sintering, respectively, demonstrating successful fabrication of the layered structure. After sintering at 1080 °C, improved densification is observed; however, the microstructure indicates that the tri-layer configuration is not fully optimized, as the three layers are not perfectly aligned with one another. Despite this, the tri-layer Al-LLZO exhibits superior electrochemical performance compared to dense Al-LLZO, suggesting that the graded architecture plays a beneficial role in enhancing cell behavior.

## 3.4. NRA (p, α)

NRA spectra were analyzed to establish the relationship between detector channel, particle energy, and lithium depth distribution. Higher channel numbers correspond to higher detected energies and near-surface lithium, whereas lower energies originate from deeper regions due to energy loss during transport. Areal densities extracted from SIMNRA (atoms cm$^{-2}$) were converted to physical thickness using the atomic number density of LLZO, enabling reconstruction of depth-resolved lithium profiles through stopping power corrections.

The normalized counts are proportional to lithium concentration, enabling comparison across samples. The experimental data were fitted using a multilayer model in SIMNRA, incorporating layer thickness and lithium concentration as fitting parameters. The tri-layer Al-LLZO exhibits a higher Li count on the surface compared to dense Al-LLZO across the energy range, indicating improved lithium transport and reduced interfacial accumulation.

Depth profiling (Fig. 4) reveals a graded lithium distribution in the tri-layer Al-LLZO, with ~75% lithium within the top ~0.59 μm, decreasing progressively to ~30% in the bulk across a total thickness of ~4.5 µm. In comparison, dense Al-LLZO displays a steeper lithium gradient, with ~48% at the surface layer (~0.35 µm) and ~16% in the bulk across a total thickness of ~2.57 µm. This contrast indicates that the tri-layer architecture promotes more effective lithium redistribution and electrochemical stability.

# 4. Conclusions

The tri-layer Al-LLZO solid state electrolyte significantly outperforms dense Al-LLZO in Li/Al-LLZO/NMC(111) full cells, delivering nearly twice the discharge capacity after 25 cycles (~55 mAh g$^{-1}$ vs. ~27 mAh g$^{-1}$). It also shows lower interfacial impedance (~373 Ω initially) and higher CE, indicating improved interfacial stability. NRA results reveal a lithium-rich surface region (~75%) in the tri-layer Al-LLZO compared to dense Al-LLZO (~48%), highlighting the role of graded microstructure in enhancing lithium transport and overall cell performance.

# 5. Acknowledgements

This research project was made possible by the generous support of the NYS Center for Advanced Technology in Nanomaterials and Nanoelectronics (CATN2). We are deeply grateful for their financial contribution and unwavering belief in our work, which has been instrumental in achieving our research goals. I would also like to thank Prof. William Lanford, Dr. Daniele Cherniak, Dr. Stephen Bedell, and Andrew Knutson for their assistance with data collection and for their valuable review of this work.

# 7. Figures

## 7.1. Figure 1

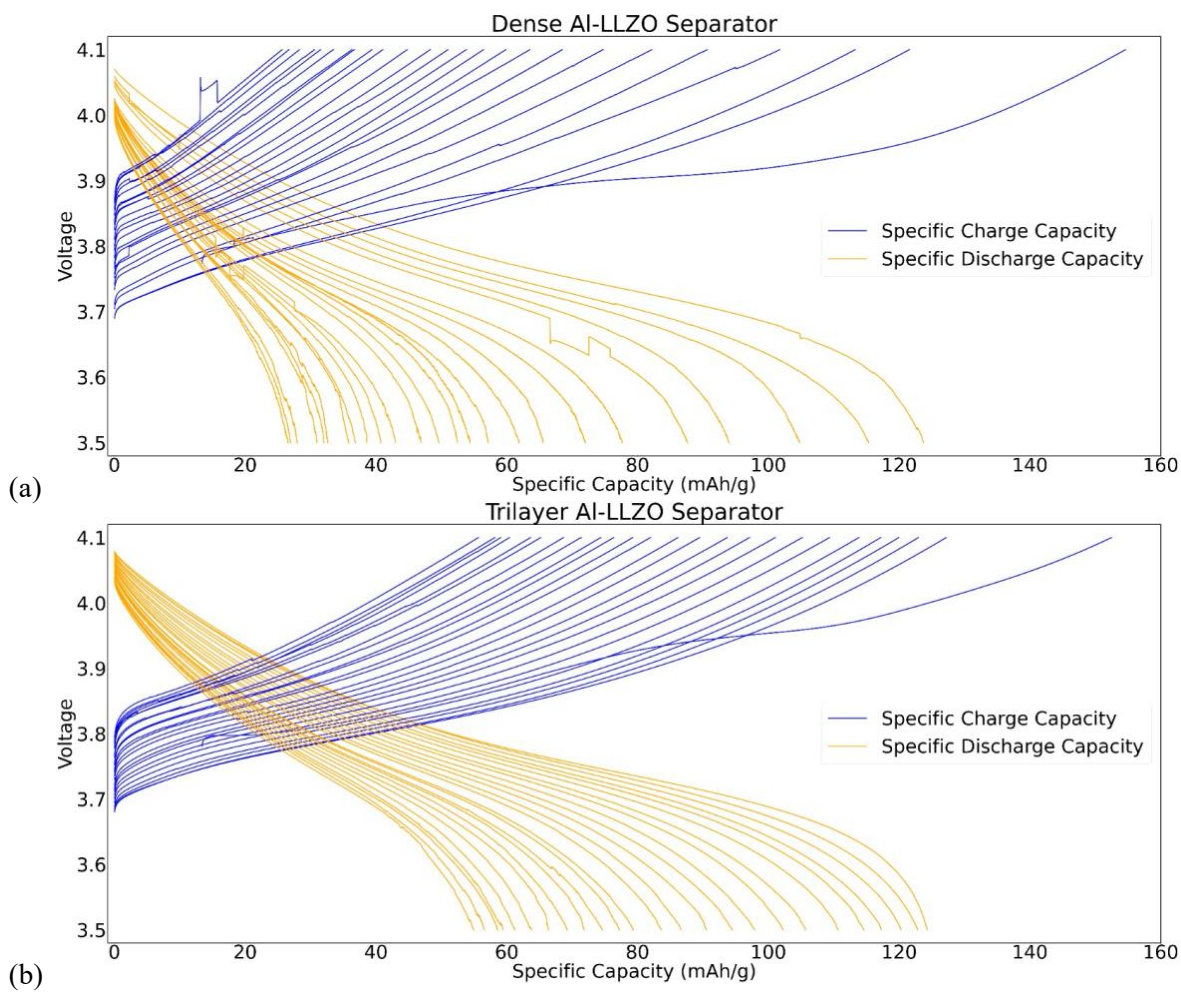

(a)

(b)

**Figure 1: Charge-discharge profile:** Cell voltage versus specific capacity for Li/Al-LLZO/NMC(111) cells using (a) dense Al-LLZO and (b) Tri-layer Al-LLZO electrolytes.

## 7.2. Figure 2

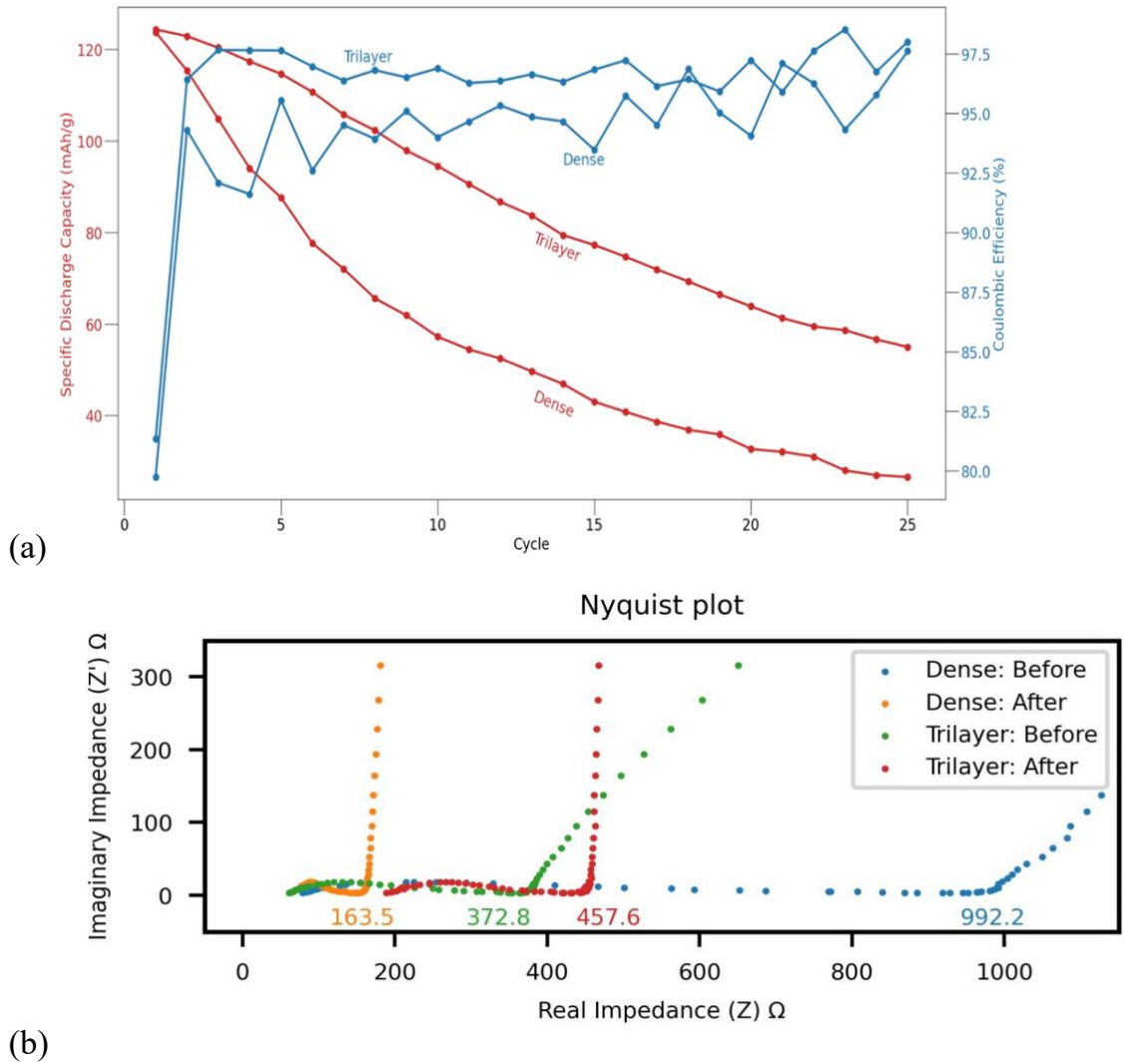

(a)

(b)

**Figure 2: Coulombic efficiency and Nyquist plot:** (a) After 25 cycles, the tri-layer Al-LLZO cell delivers a discharge capacity of 54.94 mAh g$^{-1}$, which is approximately 2 times higher than that of the dense Al-LLZO cell (26.54 mAh g$^{-1}$) and improved coulombic efficiency in tri-layer structure. (b) Nyquist plot of dense Al-LLZO cell shows a decrease in impedance after cycling, whereas the tri-layer Al-LLZO cell shows a slight increase. Impedance measurements were performed over a frequency range of 0.1 Hz to 100 kHz with an AC perturbation amplitude of 10 mv

## 7.3. Figure 3

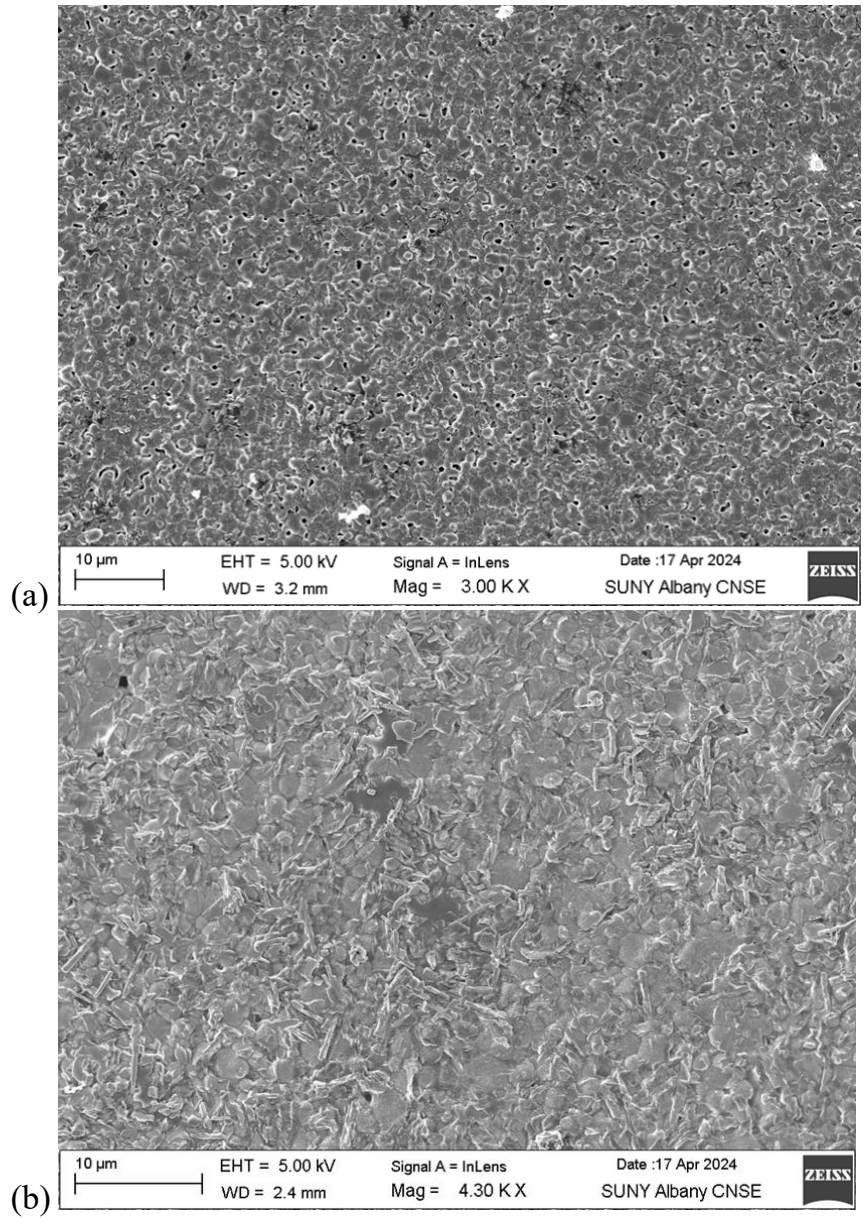
(a)
(b)

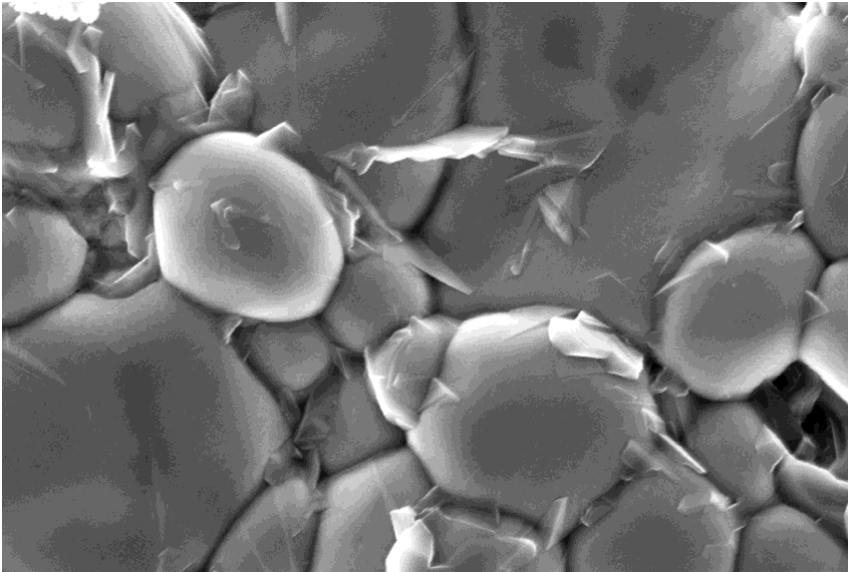

(c)

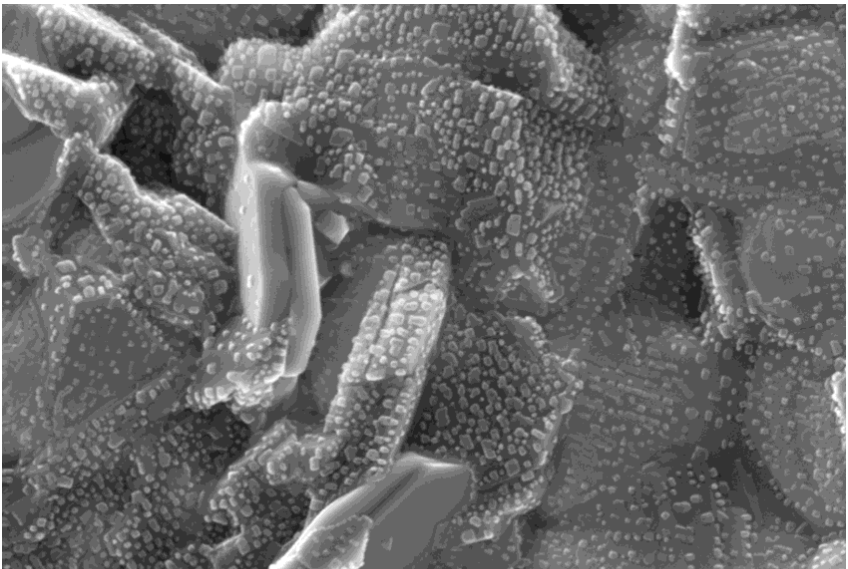

(d)

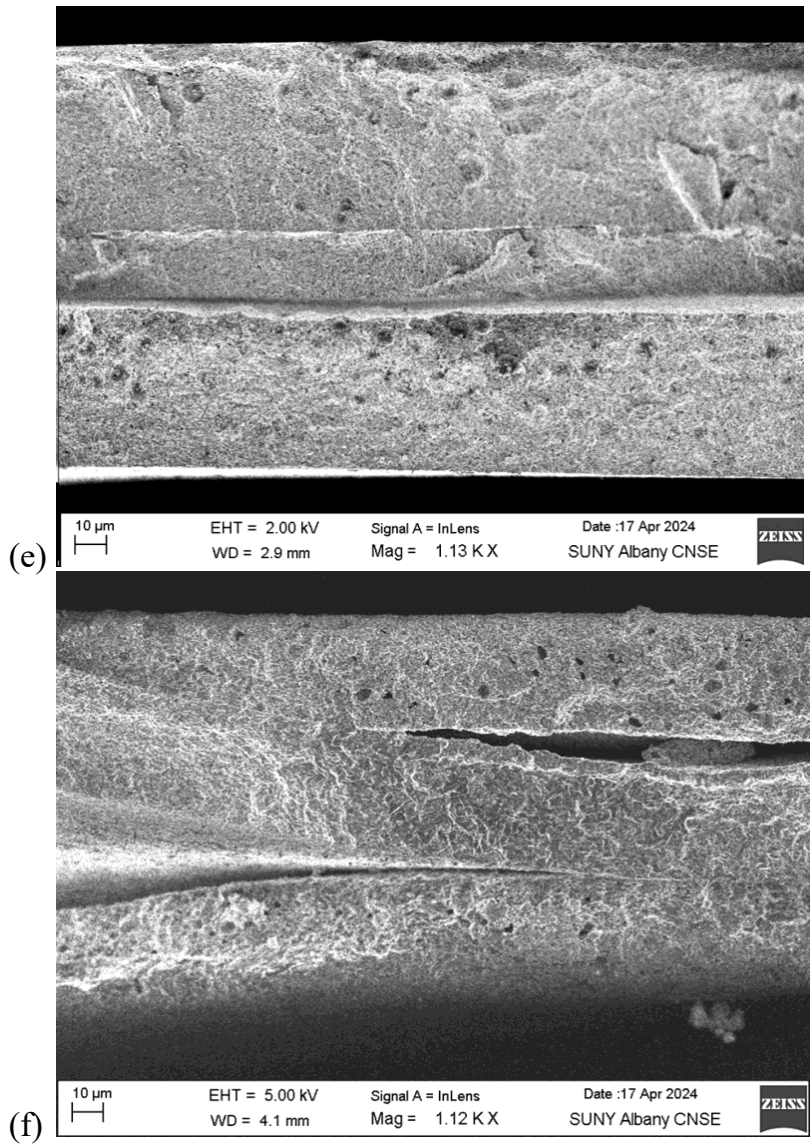

**Figure 3:SEM images :** a-d) Top view sintered pellets (a) Dense Al-LLZO (b) Porous Al-LLZO (C) Magnified @ 60,000x dense Al-LLZO (d) Magnified @ 60,000x porous Al-LLZO e&f) Cross section (e) Unsintered Al-LLZO (f) Sintered at 1080 °C Al-LLZO.

## 7.4. Figure 4

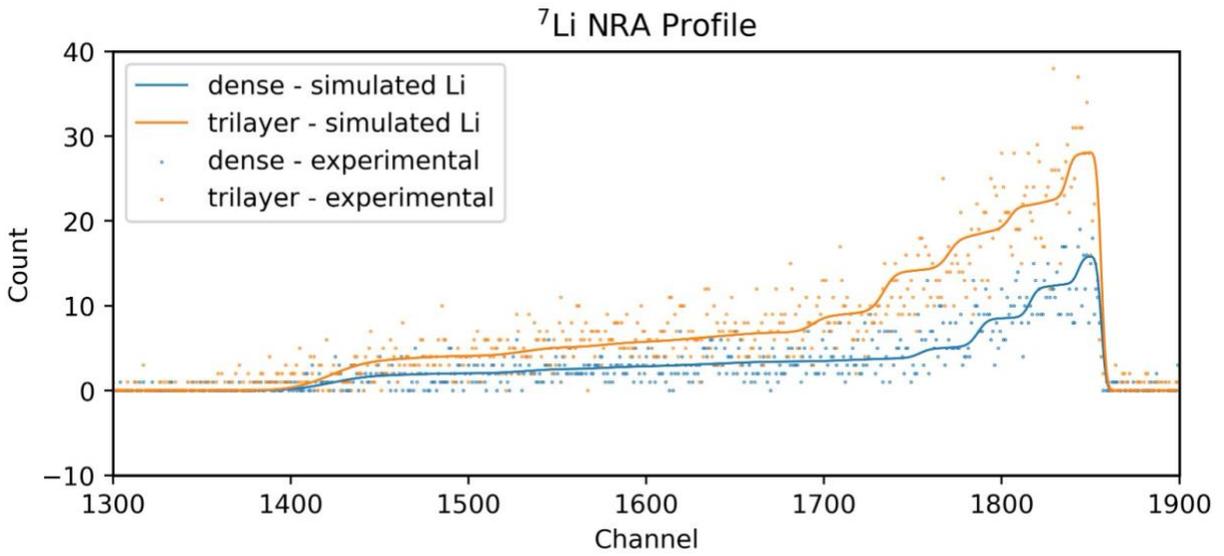

**Figure 4:NRA spectrum :** Trilayer Al-LLZO reveals higher lithium retention compared to dense Al-LLZO after 25 cycles.